\documentclass[a4paper]{spie}  

 \def\cps{counts/s}
 \def\xifu{X-IFU}
 \def\athena{{\it Athena}}
 
\usepackage{amsmath,amsfonts,amssymb}
\usepackage{graphicx}
\usepackage[colorlinks=true, allcolors=blue]{hyperref}

\title{The Athena X-ray Integral Field Unit (X-IFU)}
\author[a]{Didier Barret}
\author[b]{Thien Lam Trong}
\author[c]{Jan-Willem den Herder}
\author[d]{Luigi Piro}
\author[e]{Massimo Cappi}
\author[f]{Juhani Huovelin}
\author[g]{Richard Kelley}
\author[h]{J. Miguel Mas-Hesse}
\author[i]{Kazuhisa Mitsuda}
\author[j]{St\'ephane Paltani}
% ...
\affil[a]{IRAP, Universit\'e de Toulouse, CNRS, UPS, CNES, 9, Avenue du Colonel Roche, BP 44346, 31028 Toulouse Cedex 4, France}
\affil[b]{Centre National d'Etudes Spatiales, Centre spatial de Toulouse, 18 avenue Edouard Belin, 31401 Toulouse Cedex 9, France}
\affil[c]{SRON, Netherlands Institute for Space Research, Sorbonnelaan 2, 3584 CA Utrecht, The Netherlands}
\affil[d]{Istituto di Astrofisica e Planetologia Spaziali, Via Fosso del Cavaliere 100, 00133, Roma, Italy}
\affil[e]{INAF, Osservatorio di Astrofisica e Scienza dello Spazio, via Gobetti 93/3, 40129, Bologna, Italy }
\affil[f]{Department of Physics, Division of Geophysics and Astronomy, P.O. Box 48, FI-00014, University of Helsinki, Finland}
\affil[g]{NASA/Goddard Space Flight Center, 8800 Greenbelt Rd, Greenbelt, MD 20771, United States}
\affil[h]{Centro de Astrobiologia, CSIC/INTA, km 4, Crtra de Ajalvir, 28850, Torrejon de Ardoz, Madrid, Spain}
\affil[i]{Institute of Space and Astronautical Science (ISAS) \& Japan Aerospace Exploration Agency (JAXA), 3-1-1 Yoshinodai, Chuo-ku, Sagamihara, 252-5210, Japan}
\affil[j]{Department of Astronomy, University of Geneva, Chemin d'Ecogia 16, CH-1290 Versoix, Switzerland}
% ...

\authorinfo{Further author information: (Send correspondence to Didier Barret: dbarret@irap.omp.eu)}

\begin{document} 
\maketitle
\newpage
\begin{abstract}
The X-ray Integral Field Unit (X-IFU) is the high resolution X-ray spectrometer of the ESA \athena\ X-ray observatory. Over a field of view of 5' equivalent diameter, it will deliver X-ray spectra from 0.2 to 12 keV with a spectral resolution of 2.5 eV up to 7 keV on $\sim 5$'' pixels. The X-IFU is based on a large format array of super-conducting molybdenum-gold Transition Edge Sensors cooled at $\sim 90$ mK, each coupled with an absorber made of gold and bismuth with a pitch of 249 $\mu$m. A cryogenic anti-coincidence detector located underneath the prime TES array enables the non X-ray background to be reduced. A bath temperature of $\sim 50$ mK is obtained by a series of mechanical coolers combining 15K Pulse Tubes, 4K and 2K Joule-Thomson coolers which pre-cool a sub Kelvin cooler made of a $^3$He sorption cooler coupled with an Adiabatic Demagnetization Refrigerator. Frequency domain multiplexing enables to read out 40 pixels in one single channel. A photon interacting with an absorber leads to a current pulse, amplified by the readout electronics and whose shape is reconstructed on board to recover its energy with high accuracy. The defocusing capability offered by the \athena\ movable mirror assembly enables the X-IFU to observe the brightest X-ray sources of the sky (up to Crab-like intensities) by spreading the telescope point spread function over hundreds of pixels. Thus the X-IFU delivers low pile-up, high throughput ($>50$\%), and typically 10 eV spectral resolution at 1 Crab intensities, i.e. a factor of 10 or more better than Silicon based X-ray detectors. In this paper, the current X-IFU baseline is presented, together with an assessment of its anticipated performance in terms of spectral resolution, background, and count rate capability. The X-IFU baseline configuration will be subject to a preliminary requirement review  that is scheduled at the end of 2018. 

{\it The X-IFU will be provided by an international consortium led by France, the Netherlands and Italy, with further ESA member state contributions from Belgium, Czech Republic, Finland, Germany, Ireland, Poland, Spain, Switzerland and contributions from Japan and the United States.}
\end{abstract}

\section{Introduction}
The \xifu\ is the cryogenic micro-calorimeter of the \athena\ space X-ray observatory, the second large mission of the Cosmic Vision Program of the European Space Agency [\citen{NandraWP2013,barret_sf2a_2013,barret_arxiv_2013,ravera_spie_2014,barret_spie_2016,barcons_astnac_2017,pajot_ltd_2018}]. It is designed to 1) study  the dynamical, physical and chemical properties of hot plasmas, as those found in clusters of galaxies, and 2) study black hole accretion disks, jets, outflows and winds from galactic stellar mass black holes to the supermassive ones found in active galactic nuclei. With its unprecedented capabilities, it is also the prime instrument for many \athena\ observatory science targets, such as planets, stars, supernovae, compact objects, interstellar medium\ldots  When operated in the new era of time domain astronomy, it will also respond to target of opportunities, with triggers provided by ground and space based facilities. The \athena\ \xifu\ related scientific objectives are described in [\citen{barret_spie_2016}] (see e.g. [\citenum{cucchettia_spie_2018,roncarelli_2018}] for the most recent simulations of X-ray observations of galaxy clusters with the X-IFU). The key performance parameters of the \xifu\ are indicated in Table \ref{t_performance}. Along the redefinition of the \athena\ scientific objectives for the so-called cost-constrained mission configuration which has a 1 keV effective area reduced by 30\% and an available observing time reduced by 20\%, several scientific objectives share the same targets in the mock observing program and are now using the \xifu\ as the prime instrument. Most notably, this concerns measurements of the cluster entropy evolution with redshift and measurements of stellar mass black hole spins and winds. The former has strengthened the requirement on the X-IFU field of view as clusters at low redshifts are extended objects, while the latter is now setting the high count rate requirement, as galactic X-ray binaries are bright X-ray sources, reaching occasionally Crab-like intensities.   
\begin{table}[htp]
\caption{\xifu\ key performance requirements.}
\begin{center}
\begin{tabular}{|l|l|}
\hline
Energy range & 0.2--12.0 keV \\
Spectral resolution & 2.5 eV (up to 7.0 keV) \\
Non X-ray background & $5 \times 10^{-3} \rm ~counts/s/cm^2/keV$ (2-10 keV) \\
2.5 eV throughput (broadband, point source) & 80\% at 1 mCrab (10 mCrab as a goal) \\ 
10 eV throughput (5-8 keV, point source) & 50\% at 1 Crab \\
2.5 eV throughput (broadband, extended source) & 80\% at $2\times10^{-11}$ ergs/s/cm$^2$/arcmin$^2$ (0.2-12 keV) \\
Continuous cool time & 32 hours \\
\hline
\end{tabular}
\end{center}
\label{t_performance}
\end{table}%

\begin{figure}[!t]
\begin{center}
\includegraphics[scale=0.20]{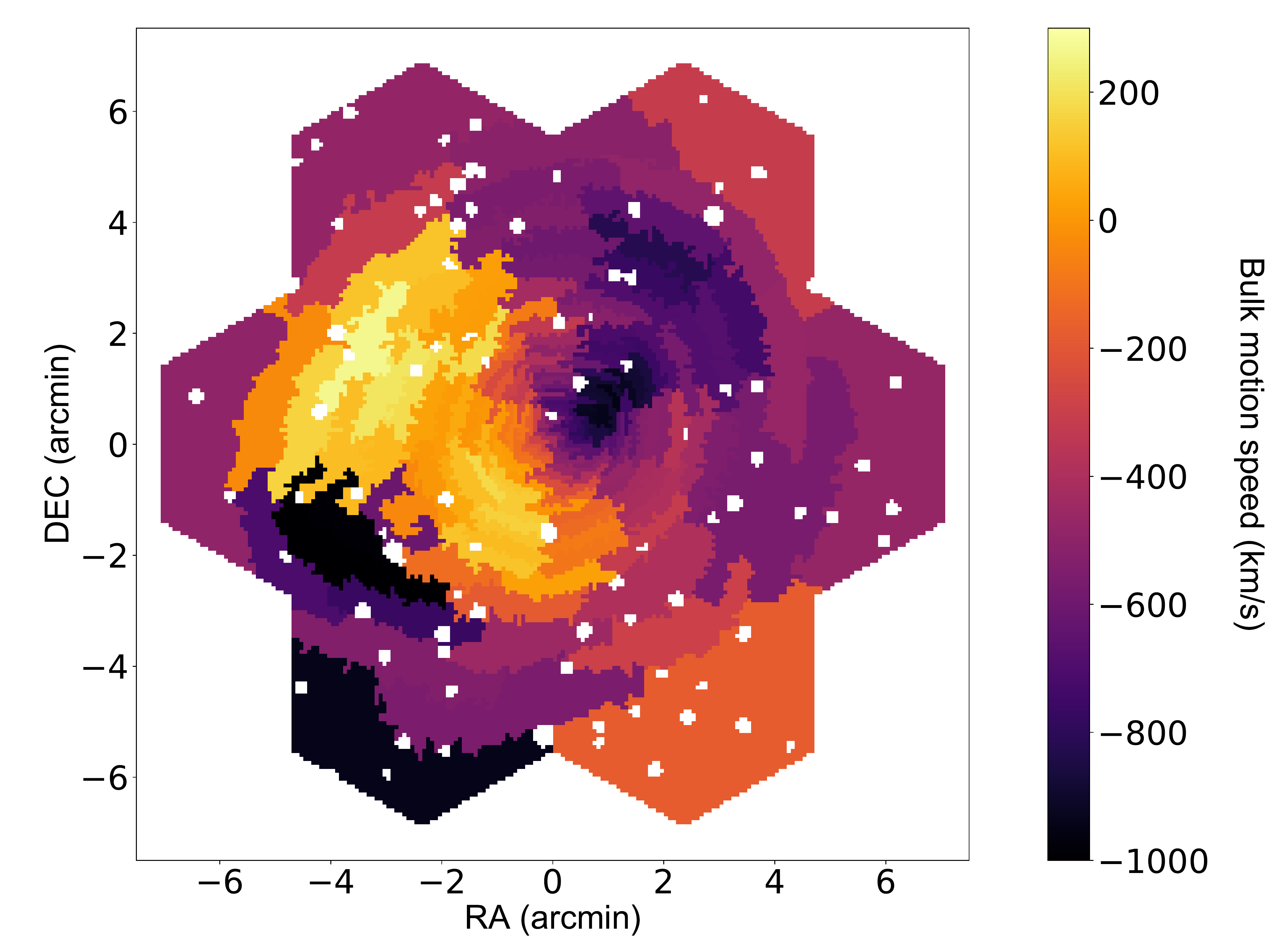}\includegraphics[scale=.20]{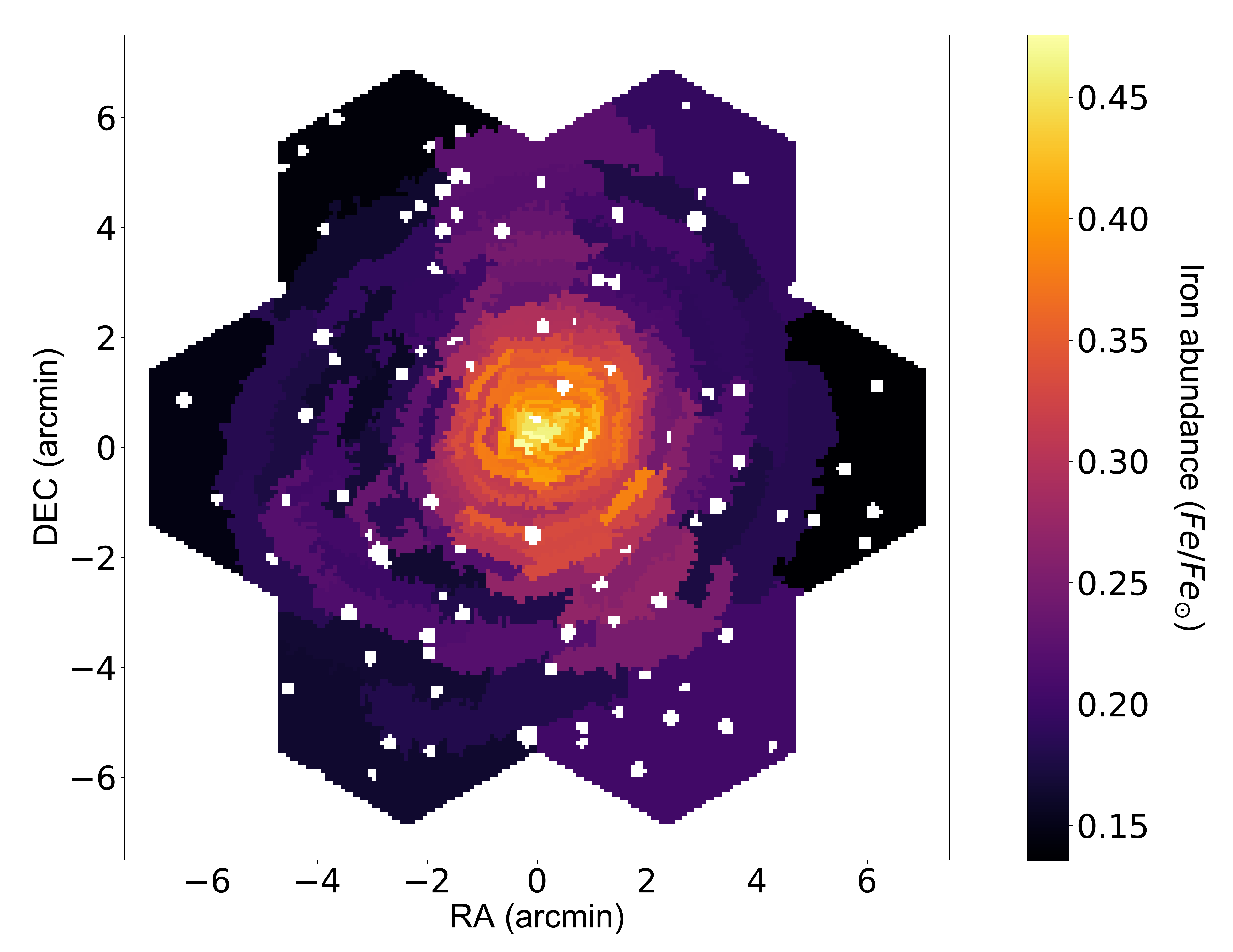}
\caption{Left) Reconstructed bulk motion velocity map (km/s). Right) Reconstructed map of iron abundance (with respect to solar). The simulations are presented in [\citen{cucchettia_spie_2018}].}
\end{center}
\end{figure}

In this paper, we will briefly describe the current instrument baseline, emphasizing on the components entering into the performance budgets. We will then discuss our current assessment of the instrument performances. This overview provides pointers to X-IFU related papers published in the same proceedings, to which the reader is referred for more detailed information. 
\section{The current \xifu\ baseline}
\subsection{Detector and focal plane assembly}
The heart of the \xifu\ is the molybdenum-gold Transition Edge Sensor (TES) array [\citen{smith_spie_2018}]. There are currently 3840 TES. The electroplated absorber is composed of $1.7\rm \mu m$ of gold and $4.2 \rm \mu m$ of bismuth, to provide the low heat capacity, the fast thermalization to the sensor and the required stopping power.  The absorber pitch is 249 $\mu m$ pitch. The 3840 absorbers define an hexagonal a field of view of 5 arcminute equivalent diameter. Optimization of the TES functional parameters is discussed in [\citen{smith_spie_2018}]. Located 1 millimeter underneath the main TES array, a cryogenic anti-coincidence TES detector is required to reduce the non X-ray instrumental background to a level of $5 \times 10^{-3}$ cts/s/cm$^{2}$/keV above 2 keV [\citen{lotti_spie_2018,dandrea_spie_2018}]. The two cryogenic detectors are supported by the Focal Plane Assembly (FPA)[\citen{jackson_spie_2018,van_weers_spie_2018}]. The FPA ensures the  thermal isolation of the detector, electromagnetic shielding and filtering while hosting the cold front end electronics of the TES (LC filters and SQUIDs). The thermal/mechanical design of the FPA development model is presented in [\citen{van_weers_spie_2018}] (see also [\citen{dandrea_spie_2018}] for the development of the cryo-AC to be integrated in the demonstration model of the FPA). 
\subsection{Cryogenic chain}
\begin{figure}[!t]
\begin{center}
\includegraphics[scale=0.45]{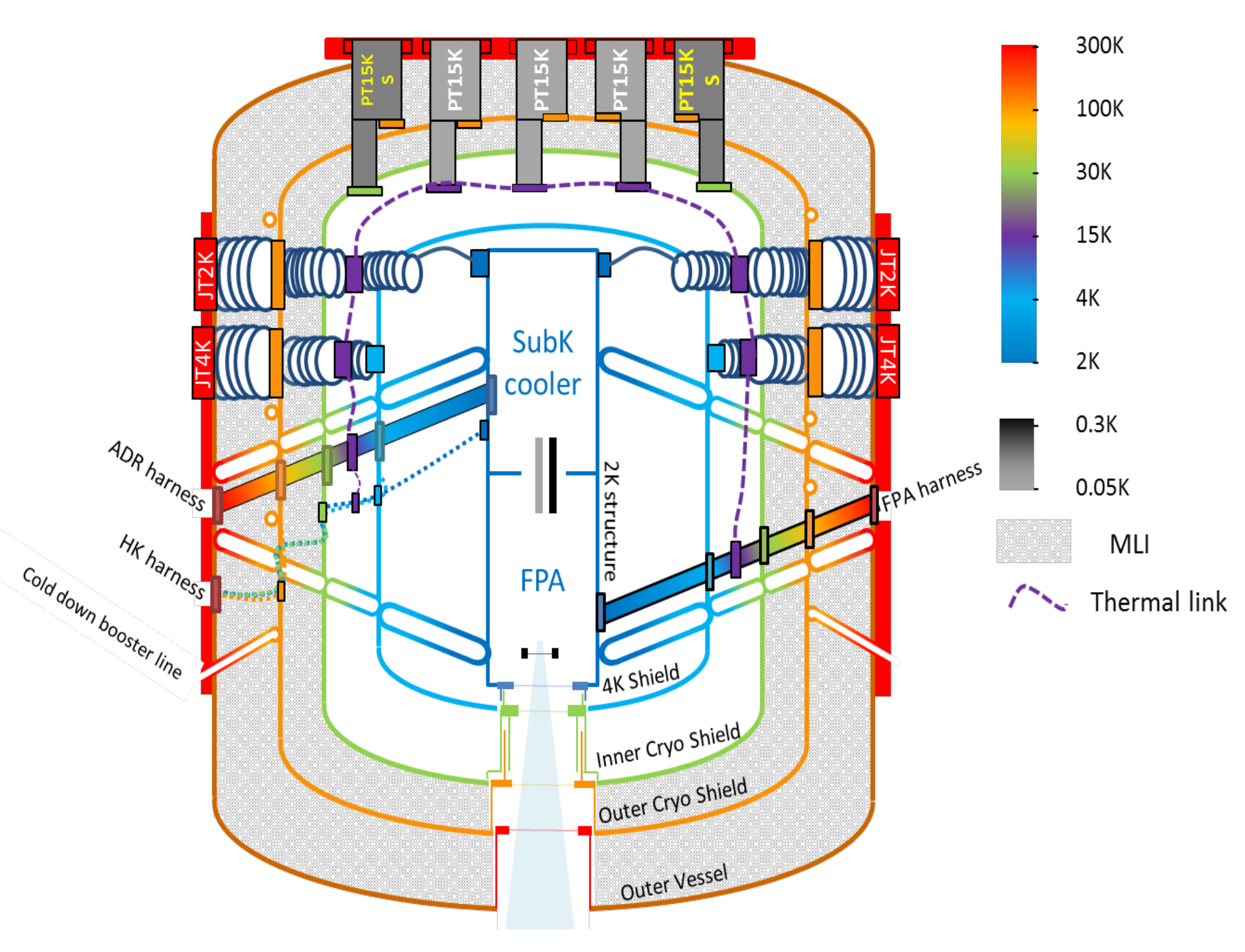}\includegraphics[scale=0.335]{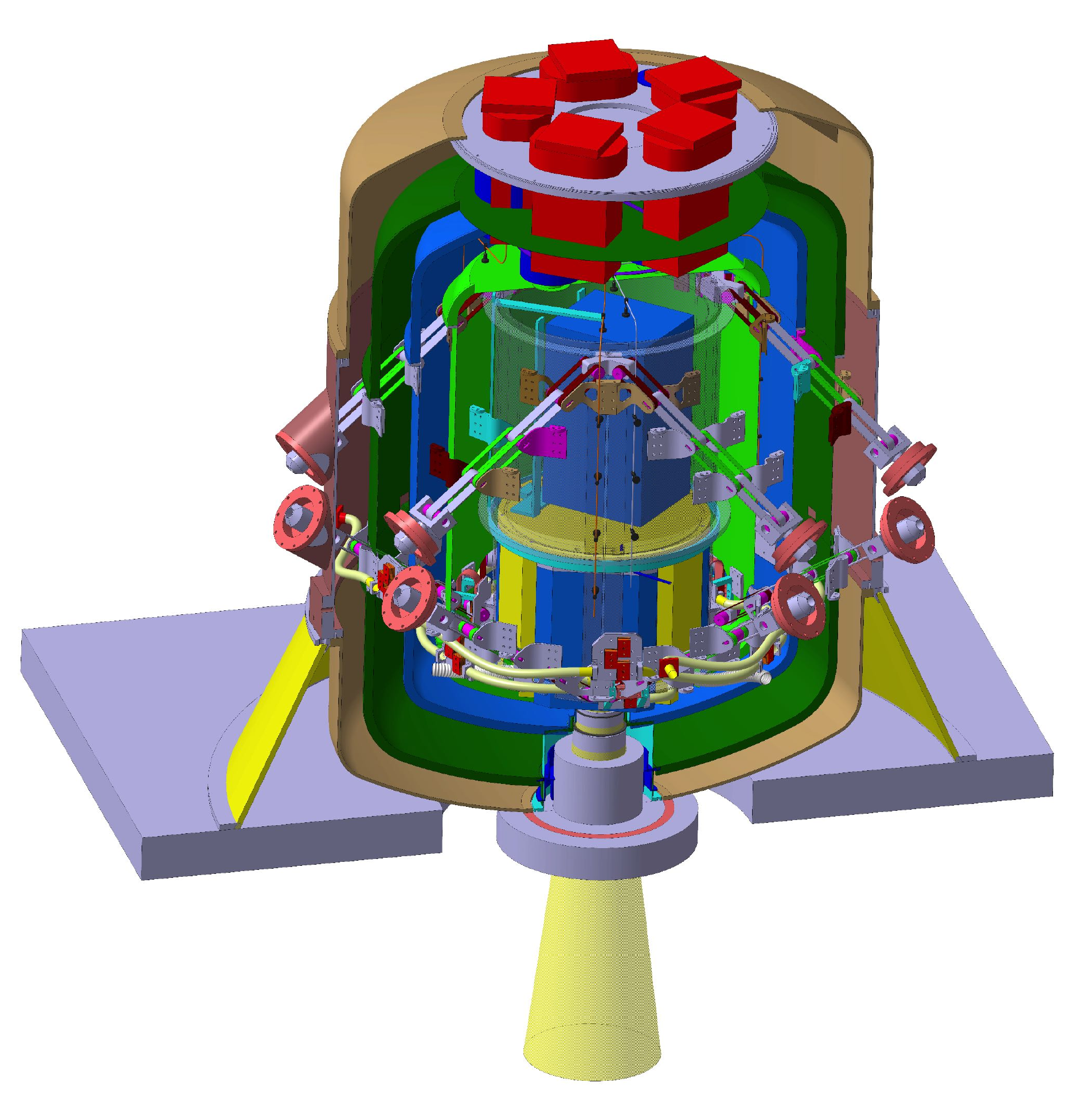}
\caption{Left) The X-IFU cryostat, emphasizing the different mechanical coolers used and the different thermal shields considered. The outer vessel has a temperature of 200K, assumed to be acheived through passive cooling. How to best achieved this interface temperature on the SIM is currently being studied, and may imply the addition of an external mechanical vessel at 300 K. Right) An open view of the X-IFU cryostat, emphasizing the 2K core and FPA (in blue) and the mechanical straps (courtesy of CNES).\label{fig:1ab}}
\end{center}
\end{figure}

The TES are operated at about 90 mK, with the bath temperature being at about 50 mK. This cold temperature is achieved by a series of mechanical coolers, associated with a multi shield cryostat (see [\citen{charles_spie_2016}] for a description of an earlier version of the X-IFU cryogenic chain architecture). The baseline cooling chain is currently composed of five 15K Pulse-Tube coolers, two 4K Joule-Thomson coolers, two 2K Joule-Thomson coolers and a last stage sorption-ADR. The cryogenic chain has been subject to an intense optimization exercise lately and is now demonstrated to achieve sufficient thermal budget margins (typically larger than 30\%), even in what is considered as a worst case scenario being the failure of a 2K Joule-Thomson cooler. The thermal budgets have been computed after revising the heat loads at the cold stage, and more importantly assuming the outer vessel of the dewar to be at about 200 K. Final refinements on the cryochain architecture are being considered, as the accommodation of the \xifu\ on the  \athena\ Science Instrument Module (SIM) is being studied in details. This is a joint effort between the ESA study team and the payload teams. The cryogenic chain is designed to achieve a cool time of 32 hours and a nominal duty cycle of 80\%. Optimization of the regeneration cycle is also investigated as to maximize the efficiency for successful completion of target of opportunity observations, requiring at least 50 kiloseconds of available cool time. The schematic of the cryostat, together with an open view on the cryostat are shown in Figure \ref{fig:1ab}. 
\subsection{Thermal and optical blocking filters}
The interface between the cryostat and the outside is achieved by the aperture cylinder, holding thermal and optical blocking filters [\citen{barbera_spie_2014,barbera_ltd_2018,barbera_spie_2018,lociciero_spie_2018,sciortino_spie_2018,parodi_spie_2018}]. The filters have to limit the radiative heat load on the detector with a contribution to the spectral resolution budget of less than 0.2 eV. They also provide radio-frequency attenuation and have to prevent molecular contamination. Five filters are currently considered, each consisting of Polyamide 45 nm thick and aluminum 30 nm thick on one side, a honeycomb supporting gold plated stainless steel mesh on four filters with the 50 mK filter made of niobium. The  blocking factor of each mesh is $\sim 3$\%. As the transmission of the filter is a key contributor to the X-IFU quantum efficiency, the blocking factor of the mesh is subject to further optimization, in an attempt to achieve blocking factors of the order of 2\%. The reduced thickness of the filters enables a higher transmission at lower energies, despite the losses due to the mesh, which penalized the response at higher energies where the filters are transparent. The thermal modeling of the filters is presented in [\citen{sciortino_spie_2018}],  the structural modeling and tests supporting their design are described in [\citen{parodi_spie_2018}], while the RF attenuation properties of thin aluminized plastic filters are discussed [\citen{lociciero_spie_2018}].
\subsection{Dewar door}
The dewar assembly will be closed by a door to protect the filters from acoustic loads during launch, to enable X-IFU measurements on the ground (and in early flight operations) and to prevent contamination during storage and operations. The door will provide an interface with the ground calibration assembly, providing sufficient transmission between calibration sources and detector (through small internal beryllium windows).
\subsection{Readout chain}
The baseline readout scheme for \xifu\ is Frequency Division Multiplexing (or Frequency Domain Multiplexing)[\citen{akamatsu_spie_2018,den_hartog_spie_2018,ravera_spie_2018,gumuchian_spie_2018,chen_spie_2018,cobo_spie_2018,peille_spie_2016}]. This enables to read 40 TES in one single channel: the whole TES array thus being read out by 96 readout channels. All elements of the readout chain have reached a high level of definition, with in some cases, prototypes already developed, such as for the Digital Readout Electronics [\citen{ravera_spie_2018,gumuchian_spie_2018}] or for the Warm Front End Electronics [\citen{chen_spie_2018}]. Testing of different triggering algorithms of the event processor is presented in [\citen{cobo_spie_2018}]. Time Division Multiplexing is also studied as a backup readout scheme for the \xifu\ [\citen{ullom_spie_2018}]. 

The spectral resolution budget which details the contribution of each component of the readout chain from the intrinsic detector noise, the amplifier noises, the bias noises, as well as the degradation from EMI/EMC and the variability of the instrument environment,  demonstrates that a 2.5 eV resolution can be achieved with a 0.5~eV (rss) margin [\citen{den_hartog_spie_2018,peillea_spie_2018}]. Recent FDM measurements of multiplexed pixels demonstrate that the required spectral resolution of 2.5 eV is within reach [\citen{akamatsu_spie_2018}]. As stated above, the TES parameters (e.g. speed) are being further optimized as described in [\citen{smith_spie_2018}], in particular in view of further relaxing the constraints on the readout electronics. This is particularly important for one of the critical elements of the readout chain, namely the Digital-to-Analogue Converters (DACs) which generate the carrier signals of the FDM and the signals of the base-band feedback loops. The DAC was identified up to now as a critical item. As demonstrated by [\citen{den_hartog_spie_2018}], the current state-of-the-art system, based on Analog Devices AD 9726 is already meeting the requirements, with an acceptable contribution to the spectral resolution budget. Space qualification of this component is being addressed now. In parallel, as part of the pixel optimization we are investigating the increase of the pixel pitch [\citen{smith_spie_2018}]. This may enable to reduce the multiplexing factor, thus providing additional margins for the design of the detector readout chain. Alternatively, this may help in reducing the number of channels, hence the heat loads on the cold stages and save mass globally at instrument level.
\subsection{Modulated X-ray sources}
A micro-calorimeter requires a careful and precise monitoring of the detector gains to correct drifts of the energy scale [\citen{cucchettib_spie_2018}]. This is achieved using modulated X-ray sources (MXS) which provide a set of very stable lines at fiducial energies. The pulsed sources considered for \xifu\ are based upon the one that is foreseen for the X-ray Imaging and Spectroscopy Mission (XRISM) Resolve instrument [\citen{kelley_spie_2016}], generating Cr-K and Cu-K lines at 5.41 keV and 8.04 keV respectively [\citen{devries_spie_2018}]. The addition of softer, e.g., Si (1.73 keV), lines is also investigated. Multi-parameter gain correction techniques and requirements on the MXS configuration are discussed in [\citen{cucchettib_spie_2018}].
\subsection{Filter wheel}
The \xifu\ filter wheel is currently baselined to have 7 positions: open, close, thin and thick optical filters, $^{55}$Fe calibration source (to be used in case of the failure of the modulated X-ray sources), a beryllium filter 100 $\mu$m, and a spare position (see [\citen{bozzo_spie_2016}] for an early definition of the filter wheel). The beryllium filter will be used when observing bright sources, to block the softer X-ray photons, helping to maximize the 10 eV throughput in the 5-8 keV range, where disk wind features are expected. The two optical filters will enable to observe optically bright X-ray sources while minimizing the optical load on the detectors, thus minimizing the spectral resolution degradation. Those are currently subject to detailed optimization, and trading a modest spectral resolution degradation against more effective area at low energies may be required. 
\subsection{Current mass and power budgets}
The current mass budget of the \xifu\ is about 710 kg (with design maturity margins), with the Dewar weighting about 210 kg, the cryogenic chain (cold fingers, compressors, cooler drive electronics) about 270 kg, and the detection chain readout and warm electronics also about 200 kg. The mass budget is still being consolidated, along with the study of the X-IFU accommodation on the Science Instrument Module. In terms of power budget, the maximum power is required during the last stage cooler regeneration and reaches about 3 kW (with design maturity margins).
\section{X-IFU current performances}
The performance assessment of \xifu\ is reported in details in [\citen{peillea_spie_2018}]. 
\subsection{Spectral resolution} The spectral resolution of the \xifu\ (TES and readout) is computed using the {\it tessim} tool (see [\citen{wilms_spie_2016,peillea_spie_2018}] for details). The spectral resolution shows little dependency with energy up to 3-4 keV and increases smoothly afterwards. It is about 2.7 eV at 10 keV. The better resolution at low energy will increase the weak narrow line sensitivity. Requirements on the knowledge of the instrument spectral resolution are discussed in [\citen{cucchettid_spie_2018}].
\begin{figure}[!t]
\begin{center}
\includegraphics[scale=1.]{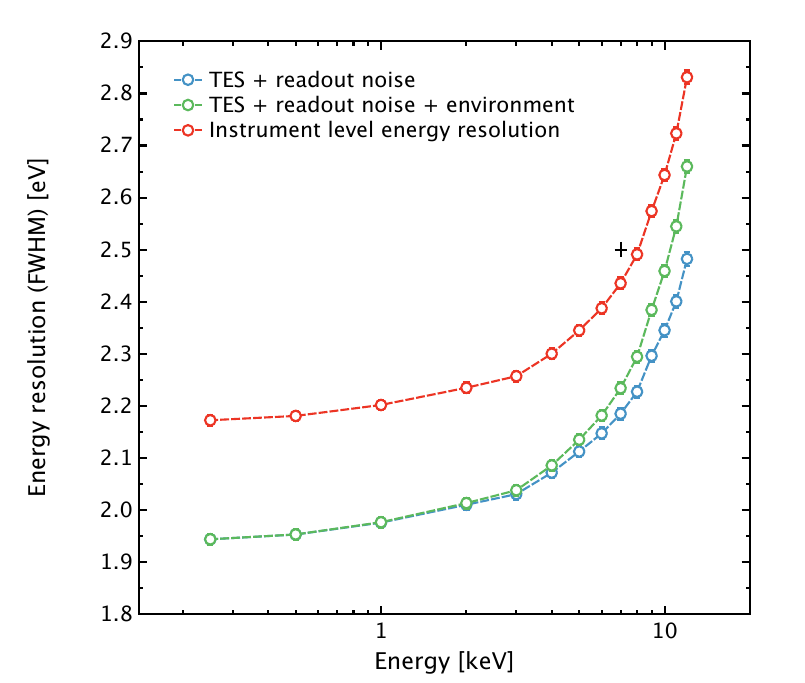}
\caption{Estimated energy resolution as a function of energy. The cross indicates the 7 keV requirement (2.5 eV). The error bars indicate  the $1\sigma$ statistical error of the simulations [\citen{peillea_spie_2018}].\label{fig:spec_res}}
\end{center}
\end{figure}
\subsection{Effective area and weak narrow line sensitivity compared to XRISM/Resolve}
\begin{figure}[!h]
\begin{center}
\includegraphics[scale=0.2]{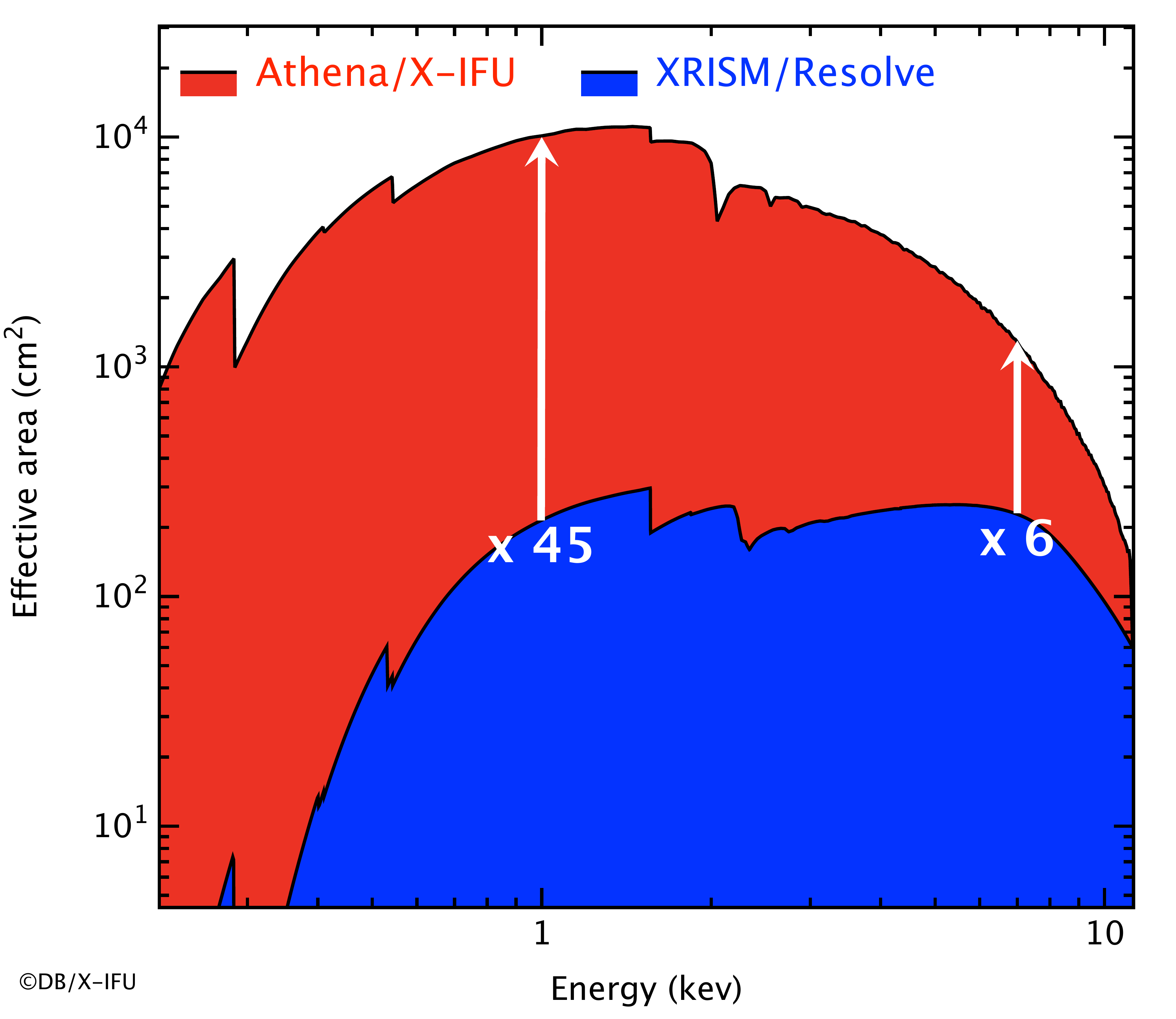}\includegraphics[scale=0.2]{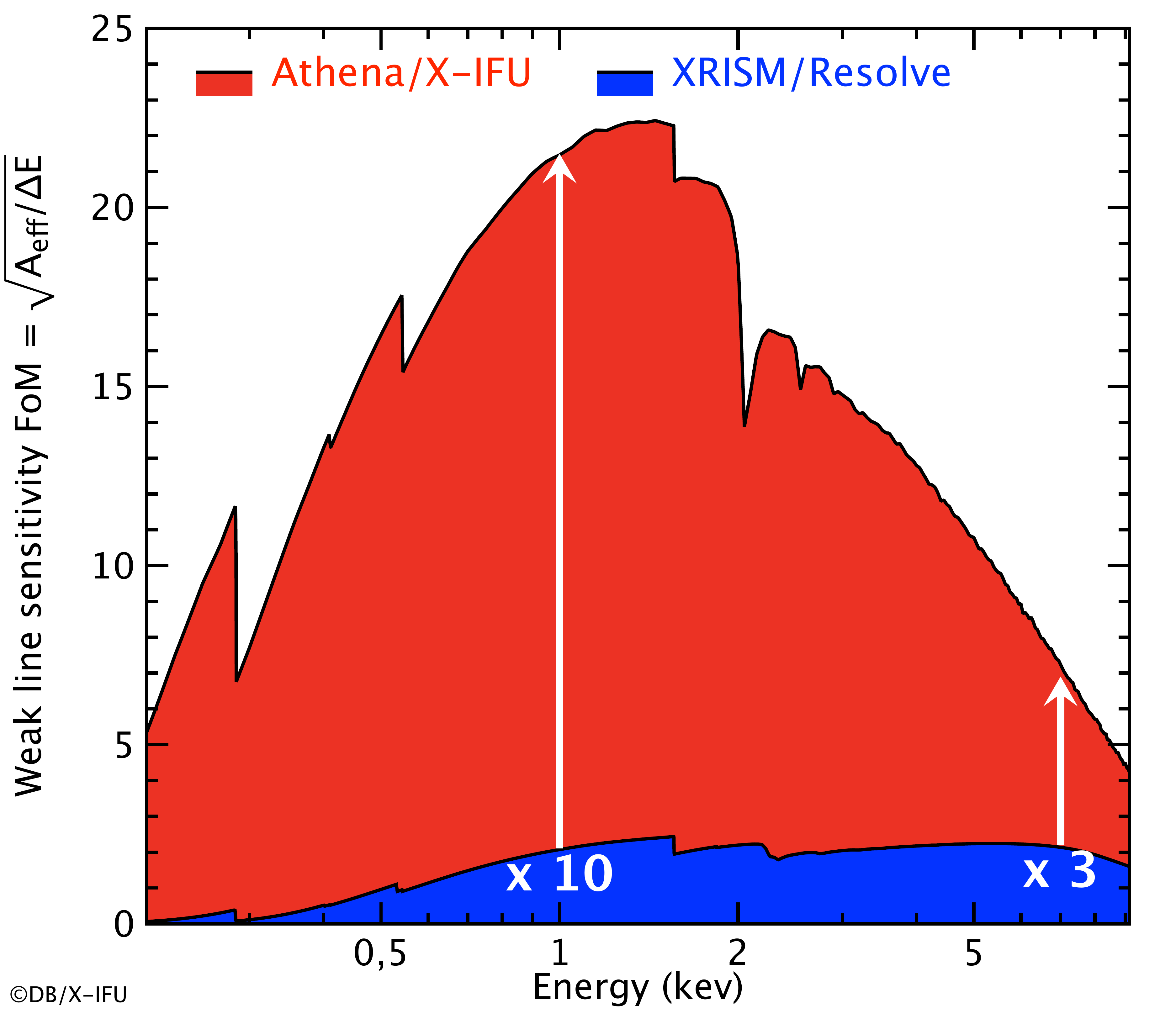}
\caption{Left) Effective area of \xifu\ in comparison with the one of XRISM/Resolve. Right) Weak line sensitivity of \xifu\ compared with XRISM/Resolve. Both figures are computed assuming the effective area of the  \athena\ optics for the cost-constrained mission. \label{f_aeff}}
\end{center}
\end{figure}

The X-IFU effective area compared to the one of the XRISM/Resolve instrument is shown in Figure \ref{f_aeff} (left) (assuming that XRISM/Resolve has the same capabilities as the Soft X-ray Spectrometer of Hitomi [\citen{kelley_spie_2016,tsujimoto_jatis_2018}]). The figure of merit for weak narrow line detection (i.e. lines with a low intrinsic width and little to no velocity/thermal broadening) scales like the square root of the effective area divided by the spectral resolution. A comparison with XRISM/Resolve is also shown in Figure \ref{f_aeff} (right). The effective area assumes a mirror effective area compliant with the  \athena\ science requirements of the cost-constrained mission. As can be seen, \xifu\ has currently a peak effective area of $\sim 1$ m$^2$ at 1 keV, i.e., $\sim 45$ times the one of XRISM/Resolve, while its weak and narrow line sensitivity will be at least an order of magnitude better at 1 keV. Requirements on the knowledge of the instrument quantum efficiency are discussed in [\citen{cucchettid_spie_2018}].

\subsection{Background}
As discussed in [\citen{lotti_spie_2018}], state of the art GEANT-4 Monte Carlo simulations indicate that the background requirement is met in the worse case scenario when the external flux of Galactic cosmic ray protons hitting the focal plane is the highest. The predicted background level is $\sim 10\%$ lower than the requirement (see Table \ref{t_performance}). The residual instrumental background remains dominated by back-scattered secondary electrons. Ways to reduce the secondary electron yield are being investigated, and different material for passively  shielding the detector considered (e.g., kapton, gold, bismuth). The use of a lateral anti-coincidence detector, surrounding the prime TES array, was investigated as a possibility to further reduce the background, but will not be baselined as the requirement is met without, and due to the increased complexity that it would bring to the design of the FPA. 

In the case of background-dominated observations of galaxy clusters outskirts, a reproducibility of 2\% on the absolute knowledge of the background is required for measuring abundances and turbulence. Different ways of monitoring the non X-ray background in-flight, including the use of the Galactic Cosmic ray background spectrum seen by the cryogenic anti-coincidence detector are investigated in [\citen{cucchettic_spie_2018}].

\subsection{Count rate}
\begin{figure}[!t]
\begin{center}
\includegraphics[scale=1.]{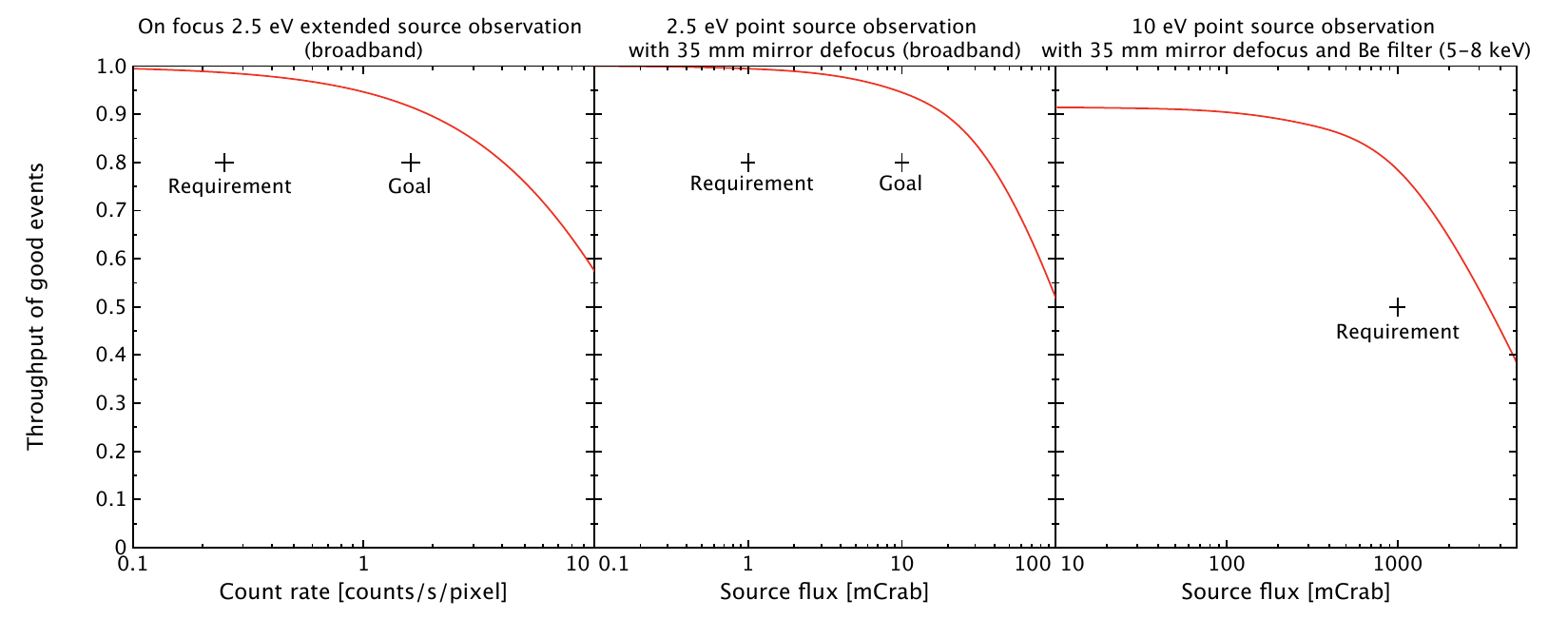}
\caption{Throughput of good events for the three sizing science cases for the X-IFU count rate capability. Left) 2.5~eV throughput for the observation of an extended source with a thermal spectrum. Middle) 2.5 eV throughput for the observation of a bright defocussed Crab-like point source. 10 mCrab corresponds to a count rate of $\sim 650$ cps over the array spread over more than 600 pixels, seeing at most 1.2 \cps. Right) 10 eV throughput for the observation of a bright defocussed Crab-like point source with a beryllium filter, removing most of softer photons below 3 keV. In all plots, crosses indicate the X-IFU count rate requirements and goals. Cross talk is accounted for as a throughput loss [\citen{peillea_spie_2018}].\label{f_timing}}
\end{center}
\end{figure}

Thanks to the defocusing capability of the \athena\ optics, its point spread function can be spread over a large number of pixels. Defocusing depth of up to 35 millimeters are considered, and this makes the \xifu\ a very powerful timing instrument, up to very bright Crab-like intensity sources [\citen{peille_ltd_2018}]. The count rate capability of the \xifu\ depends on the pixel speed, the record length required to achieve the spectral resolution, and the crosstalk level ([\citen{peillea_spie_2018}] for detailed information). As can be seen from  Figure \ref{f_timing} the requirements and the goals as stated in Table \ref{t_performance} are exceeded in all cases. This motivates the reduction of the pixel speed to make them easier to read (the decay time of the pulse is currently $\sim 300\mu$s, a value that can be further increased). Alternatively, this also gives margins to make the pixels larger: this is part of the TES optimization exercise.

\section{Conclusions}
The \xifu\ has reached a stable baseline configuration, in particular with its cryogenic chain, which demonstrates required thermal margins. The key performance requirements are met, in some case with margins, with the current baseline. Detailed accommodation of the \xifu\ on the science instrument module is being studied, while keeping the mass and power budgets within acceptance. A final round of instrument optimization is currently being performed with the goal of consolidating further the baseline in view of the instrument preliminary requirement review to be held at the end of 2018. 

Hitomi has clearly opened the window of spatially resolved high-resolution spectroscopy (see e.g. [\citen{hitomi_d,hitomi_c,hitomi_b,hitomi_a}] for a series of transformational papers describing the wealth of the Perseus observations), and it can be anticipated that after XRISM flies, X-ray astronomy will be a completely new and different field. This is one of the key reasons to keep designing the \xifu\ as a very ambitious instrument to meet its challenging requirements within resource allocations. Strong emphasis is being put on optimizing the spectral resolution, field of view and count rate capability. While we are completing the feasibility study of the X-IFU, it is remarkable to note that those key performance requirements as originally specified can be met with current technologies.
\section{Acknowledgements}
We are grateful to all the members of the X-IFU consortium for their dedication and support to the building of the instrument from science related activities to developing enabling technologies. We also wish to thank the ESA study team for their support throughout the feasibility study phase of the X-IFU.

Part of this work has been funded by the Spanish Ministry MINEICO under project ESP2016-76683-C3-1-R, co-funded by FEDER funds.
\bibliographystyle{spiebib} 
\bibliography{db_ref_spie2018} 

\end{document}